\documentclass[twocolumn,groupedaddress,amsmath,amssymb,a4paper,twoside]{revtex4}
\newcommand{\pagenumbaa}{1}
\usepackage{graphicx}
\usepackage[hidelinks=true]{hyperref}   

\newcommand{\braket}[2]{\mbox{$ \langle #1 | #2 \rangle $}}

\newcommand{\ket}[1]{\mbox{$ | #1 \rangle $}}
\newcommand{\bra}[1]{\mbox{$ \langle #1 | $}}

\begin{document}

\title{Tight steering inequalities from generalized entropic uncertainty relations}

\author{Tam\'as Kriv\'achy}
\author{Florian Fr\"owis}
\author{Nicolas Brunner}
\affiliation
{Department of Applied Physics, University of Geneva, 
CH-1211 Geneva, Switzerland}

\begin{abstract}
We establish a general connection between entropic uncertainty relations, Einstein-Podolsky-Rosen steering, and joint measurability. Specifically, we construct steering inequalities from any entropic uncertainty relation, given that the latter satisfies two natural properties. We obtain steering inequalities based on R\'enyi entropies. These turn out to be tight in many scenarios, using max- and min-entropy. Considering steering tests with two noisy measurements, our inequalities exactly recover the noise threshold for steerability. This is the case for any pair of qubit 2-outcome measurements, as well as for pairs of mutually unbiased bases in any dimension. This shows that easy-to-evaluate quantities such as entropy can optimally witness steering, despite the fact that they are coarse-grained representations of the underlying statistics.
\end{abstract}

\maketitle

\setcounter{page}{\pagenumbaa}
\thispagestyle{plain}


\section{Introduction}
\label{sec:introduction}

One of the most prominent contributions of quantum theory to our comprehension of reality is that there exist inherently unpredictable properties of certain systems. A natural quantifier of such unpredictability is entropy, in terms of which the uncertainty principle is concisely expressed as an entropic uncertainty relation (EUR) \cite{maassen_generalized_1988,coles_entropic_2017}
\begin{equation}
\label{eq:MUUR}
H(X) + H(Z) \geq q(X,Z),
\end{equation}
where $X$ and $Z$ are some observables we use to probe the system at hand, and $H(\cdot)$ denotes the Shannon entropy of the measurement statistics. The lower bound is state-independent and given by $q(X,Z) = -\log(c_{X,Z}^2)$, where $c_{X,Z} $ is the maximum absolute overlap of the eigenvectors of $X$ and $Z$.

Most astoundingly, EURs can be broken through the use of quantum correlations \cite{berta_uncertainty_2010,renes_conjectured_2009}. 
Consider two separate observers, Alice and Bob, sharing a quantum state $\rho_{AB}$. Bob measures observables $X_B$ and $Z_B$ on his subsystem, whereas Alice measures $X_A$ and $Z_A$. Bob will now test the EUR, and he can use side-information provided by Alice, e.g.~the result of the measurement she performs on her subsystem. If $\rho_{AB}$ is a separable state, i.e.~featuring only classical correlations, then Alice holds classical side-information on Bob's system. Intuitively, the latter should not change the fundamental character of quantum uncertainty. For Shannon entropies, one can indeed prove that the conditional EUR holds \cite{berta_uncertainty_2010}
\begin{equation}
H(X_B|X_A) + H(Z_B|Z_A) \geq q(X_B,Z_B).
\label{eq:shannon-entr-ineq}
\end{equation}
The situation is however completely different when Alice and Bob share entanglement. Notably, the above inequality can be violated, as Alice now provides genuinely quantum side-information \footnote{In fact, the lower bound of the inequality now becomes $q(X_B,Z_B) + H(B|A)$, where $H(B|A)$ is the conditional von Neumann entropy of $\rho_{AB}$ which can become negative when $\rho_{AB}$ is entangled \cite{berta_uncertainty_2010}.}. 

It then follows that the observation of a violation of the conditional EUR \eqref{eq:shannon-entr-ineq} implies the presence of genuinely quantum correlations between Alice and Bob. Moreover, since the measurements of Alice are not specified (only their results, but not what the observables actually are), the EUR \eqref{eq:shannon-entr-ineq} can be viewed as a steering inequality \cite{walborn_revealing_2011,schneeloch_einstein-podolsky-rosen_2013}. Although there exist other methods tailored to the detection of steering \cite{cavalcanti_quantum_2017}, entropic inequalities are appealing due to their conceptual simplicity. A violation of this entropic inequality implies the presence of EPR steering, a form of quantum correlations intermediate between entanglement and Bell nonlocality \cite{wiseman_steering_2007,quintino_inequivalence_2015} and a resource for quantum information processing \cite{branciard_one-sided_2012,piani_necessary_2015}. 

From the above statements, it is clear that EUR and steering are directly connected. A deep understanding of this connection is, however, still missing. Although steering inequalities could be derived for Shannon entropies \cite{walborn_revealing_2011,schneeloch_einstein-podolsky-rosen_2013,karthik_joint_2015,riccardi_multipartite_2018}, and more recently for Tsallis entropies \cite{costa_steering_2017}, it is not known whether a nontrivial steering criterion can be obtained from a general EUR, e.g.~based on R\'enyi entropies. Another natural question is whether EURs could provide necessary and sufficient conditions for steering.  

Here we investigate these questions and establish a general connection between EUR and steering. We develop a simple method for constructing a steering inequality from any EUR, given that the latter satisfies two natural properties. Basically, we demand that (i) the EUR holds when conditioned on classical side-information, and (ii) the entropy used in the EUR should not increase when conditioning on additional information. The simplicity of these requirements allows us to apply the connection to R\'enyi EURs, resulting in R\'enyi steering inequalities. Next, we show that some of these inequalities, based on the operationally relevant min- and max-entropies \cite{konig_operational_2009}, optimally detect steering. Specifically, when Alice and Bob share a pure entangled state (of arbitrary dimension $d \times d$), and Alice uses two mutually unbiased measurements with independent white noise, our R\'enyi steering criteria are necessary and sufficient for steering. Moreover, for $d=2$, this extends to all two-outcome quantum measurements. Finally, our paper also connects EURs to another notion of the incompatibility of quantum measurements, namely joint measurability \cite{busch_quantum_1996,heinosaari_invitation_2016}, which is itself directly connected to steering \cite{quintino_joint_2014, uola_joint_2014,uola_one--one_2015}.

\section{Preliminaries}
\label{sec:steering}

In the task formulation of the steering scenario \cite{wiseman_steering_2007,cavalcanti_quantum_2017} an untrusted party, Alice, distributes a quantum system to a trusted party, Bob. Alice tries to convince Bob that they share an entangled quantum state. To do so, Bob can choose a measurement setting (in each round of the game), and ask Alice to remotely ``steer'' the state of his subsystem in that basis. For instance, consider the case where Alice and Bob share a two-qubit singlet state. Upon receiving Bob's basis choice, Alice can perform a measurement on her subsystem in the same basis, the result of which allows her to perfectly guess Bob's measurement outcome. She then communicates her guess to Bob, who can check the perfect anti-correlation and then conclude that his subsystem is indeed entangled with Alice's. 

More generally, Bob's task is to determine whether the statistics of his measurements, conditioned on some information communicated by Alice, necessarily implies the presence of entanglement, or whether it could have actually been obtained without using any entanglement at all. To do so, Bob must check that any possible strategy in which Alice uses a non-entangled state is incompatible with the observed data. If this is the case, he will be convinced, and we say that Alice can steer Bob. If not, then the observed data admits a local hidden state model, and there is no steering. 

More precisely, any local hidden state model involves two ingredients. First, a shared classical-quantum (hence non-entangled) state 
\begin{equation}
\label{eq:clqustate}
\rho_{AB} = \sum_{\lambda} p(\lambda) \ket{\lambda} \bra{\lambda}_A \otimes \sigma^{\lambda}_B ,
\end{equation} 
where the states $\ket{\lambda}$ form an orthonormal basis, and $ \sum_{\lambda} p(\lambda) = 1$. Here the variable $\Lambda$ (with realizations $\lambda$) can be seen as a classical memory for Alice describing the quantum state $\sigma^{\lambda}_B$ that she sends to Bob. Second, a local response function for Alice which determines the classical information she will communicate to Bob depending on his measurement choice, as well as her classical memory $\Lambda$. 

In the next section, we will show how to obtain steering criteria starting from EURs. After that, we will discuss illustrative examples involving a pair of general quantum measurements for Alice, namely positive operator valued measures (POVMs), denoted $X_A$ and $Z_A$. Each POVM is a set of Hermitian positive semidefinite operators that sums to identity. We have that $X_A = \{X_A^x\}_x$ such that $X_A^x\geq0$ and $\sum_x X_A^x = \mathbb{I}$, and similarly for $Z_A$. 

In general the observation of steering from Alice to Bob requires Alice's measurements to be incompatible. The natural notion of incompatibility here is joint measurability \cite{busch_quantum_1996,heinosaari_invitation_2016}. We say that $\{X_A^x\}_x$ and $\{Z_A^z\}_z$ are jointly measurable if there exists a parent POVM $\{G^{xz}\}_{xz}$ such that $\forall x: \, X_A^x = \sum_z G^{xz}$ and $\forall z: \, Z_A^z = \sum_x G^{xz}$. Operationally, joint measurability corresponds to the property that given any input state, the statistics of $X_A$ and $Z_A$ can be obtained simultaneously, by first measuring $\{G^{xz}\}_{xz}$ and then applying classical post-processing. It turns out that the connection between steering and joint measurability is even stronger. Any set of POVMs that is not jointly measurable can give rise to steering, when Alice and Bob share a pure entangled state \cite{quintino_joint_2014,uola_joint_2014,uola_one--one_2015}. 

\section{General connection of uncertainty relations and steering}
\label{sec:generalURsteering}

In order to derive an entropic steering inequality from an EUR based on an entropy $\tilde{H}$, with a state-independent bound $\tilde{q}$ such as $\tilde{H}(X)+\tilde{H}(Z)\geq \tilde{q}(X,Z)$, we show that the following two conditions are sufficient.
\begin{itemize}
\item[(i)] \emph{Conditional EUR} The considered EUR holds true when conditioned on any classical random variable $Y$, 
\[\tilde H(X|Y) + \tilde H(Z|Y) \geq \tilde q(X,Z).\]
\item[(ii)] \emph{Conditioning Reduces Entropy} The considered entropy must not increase under conditioning on additional information, that is, for any random variables $X,Y_1,Y_2$ 
\[\tilde H(X|Y_1)\geq \tilde H(X|Y_1,Y_2).\]
\end{itemize}
Both conditions are quite natural and are expected to be satisfied for commonly used entropies. Condition (ii), often referred to as the ``conditioning reduces entropy'' property of an entropy \cite{iwamoto_information_2014}, is in general weaker than the celebrated data processing inequality, which, in turn, is usually a prerequisite for any measure of uncertainty to be used as an entropy. Condition (i) is intimately connected to the concavity of the entropy $\tilde{H}$. In Appendix \ref{app:weaker_conditions}, we give an alternative derivation of the result starting from the concavity of the conditional entropy. 

To see why conditions (i) and (ii) imply a steering criteria, let us assume a local hidden state model where Alice has classical side-information $\Lambda$ as well as a response function, with which she can determine the classical information she will communicate to Bob, i.e.~$X_A$ and $Z_A$. 
If Bob had access to this information then condition (i) would imply, for $Y\equiv (X_A,Z_A)$,
\begin{equation}
\label{eq:genericentropy_xaza}
\tilde H(X_B|X_A,Z_A) + \tilde H(Z_B|X_A,Z_A) \geq \tilde q(X_B,Z_B).
\end{equation}
However, in a local hidden state model Bob does not have access to all this information, since in each round he only asks Alice to perform one measurement and give the corresponding result. When Bob performs measurement $X_B$ ($Z_B$), Alice communicates to him the result of her measurement $X_A$ ($Z_A$). Formally, one can use condition (ii) term-wise on inequality (\ref{eq:genericentropy_xaza}) to leave out part of the information that Bob does not acquire, and arrive at
\begin{equation}
\tilde H(X_B|X_A) + \tilde H(Z_B|Z_A) \geq \tilde q(X_B,Z_B).
\label{eq:ineqgeneric}
\end{equation}
A violation of this inequality excludes the existence of a local hidden state model. Thus (\ref{eq:ineqgeneric}) is a steering inequality, whose violation is in general a sufficient, but not necessary condition for steerability.

Not only does the above argument generalize to any EUR with a state-independent bound, but also allows us to establish a one-to-one mapping of entropic steering inequalities and EUR, as 
\begin{align*}
\label{eq:one-to-one}
\sum_k \tilde{H}_k (X_{B,k}) &\geq \tilde q(\{X_{B,k}\}_k) \,\, \text{is an EUR.}\\
&\Updownarrow \text{(i) \& (ii)}\\
\sum_k \tilde{H}_k (X_{B,k} | X_{A,k}) &\geq \tilde q(\{X_{B,k}\}_k) \,\, \text{is a steering ineq.}
\end{align*}
Here the EUR relates multiple measurements $X_{B,k}$, each evaluated by a different entropy, denoted by $\tilde{H}_k$, each of which must comply with condition (ii). In the steering criteria, the $X_{A,k}$ are the corresponding responses of Alice.

Going from the EUR to the steering criteria is a straightforward generalization of previous arguments, and relies on (i) and (ii). The other direction is simple once one realizes that the steering inequality must hold for any local hidden state model. In particular it holds for \emph{any} state given to Bob and for \emph{any} measurements he performs. Thus by applying condition (ii) term-wise on the left-hand side to leave out the $X_{A,k}$ measurement statistics, one arrives at an equation of the desired form, which holds for any state and any measurement, hence it is truly an EUR.

We note that the procedure only works for state-independent bounds. For a state-dependent bound, Bob would have to have a description of his conditional state $\rho_{B|\Lambda}$, i.e.~his local hidden state, in order to evaluate the steering criteria resulting from this mechanism. However, Bob does not have access to this.

On a final note, let us point out how the derivation is explicitly connected with the joint measurability of Alice's observables. In Eq.~(\ref{eq:genericentropy_xaza}) we substituted the classical random variable $Y\equiv(X_A,Z_A)$ into condition (i) for both entropy terms. This we can only do if the results of $X_A$ and $Z_A$ are jointly acquirable in each round of the game. 

\section{R\'enyi steering inequalities}
\label{sec:renyiSI}

The above method can be applied to a wide range of EURs, see e.g. Ref.~\cite{coles_entropic_2017,schwonnek_additivity_2018}. For instance, starting from the Maassen--Uffink EUR (\ref{eq:MUUR}) we recover well-known entropic steering inequalities \cite{walborn_revealing_2011,schneeloch_einstein-podolsky-rosen_2013}. More importantly, our method allows us to construct steering inequalities based on R\'enyi entropies, for which previous approaches failed. Notably, our R\'enyi steering inequalities turn out to be tight in certain natural cases, thus outperforming all previous entropic steering criteria.

We use the conditional R\'enyi entropy of order $\alpha\geq 0$, defined as \cite{arimoto_information_1977,iwamoto_information_2014}
\begin{equation}
	H_\alpha(X|Y)=\frac{\alpha}{1-\alpha} \log \left[
\sum_y p_Y(y)
\|\mathbf{p}_{X|Y}(\mathbf{x}|y)\|_\alpha
\right],
	\label{eq:def-renyi-cond}
\end{equation}
where $\|\cdot\|_\alpha$ denotes the $\alpha$-norm and $\mathbf{p}_{X|Y}(\mathbf{x}|y)$ is a vector of conditional probabilities $p_{X|Y}(x|y)$ for a fixed $y$. For this entropy, we have the following EUR
\begin{equation}
\label{eq:RenyiEUR}
H_\alpha(X) + H_\beta(Z) \geq q(X,Z),
\end{equation}
for any $\alpha,\beta \geq 1/2$ such that $\alpha^{-1}+\beta^{-1}=2$ \cite{maassen_generalized_1988}. Condition (ii) was shown in Ref.~\cite{iwamoto_information_2014}, while condition (i) comes from the weakening of the tripartite EUR with quantum side-information \cite{tomamichel_uncertainty_2011,coles_uncertainty_2012,muller-lennert_quantum_2013-1}.
In such a tripartite scenario, Bob, Charlie, and Diana share a joint quantum state $\rho_{BCD}$ (in the steering scenario imagine that Alice holds subsystems $C$ and $D$). Either Charlie or Diana sends their state to Bob, who then performs either measurement $X$ on $\rho_B$ if he received subsystem $C$, or $Z$ on $\rho_B$ if he received $D$. Then the uncertainties of his outcomes, while having access to quantum memories $C$ or $D$, in terms of conditional R\'enyi entropies, are bounded by
\begin{equation}
\label{eq:tripartite}
H^Q_\alpha(X|C) + H^Q_\beta(Z|D) \geq q(X,Z),
\end{equation}
where $H^Q_{\alpha,\beta}$ are now quantum entropies \cite{muller-lennert_quantum_2013-1,wilde_strong_2014} written for the post-measurement reduced classical-quantum states $\rho_{XC}$ and $\rho_{ZD}$. The only property of quantum conditional R\'enyi entropies that is relevant for our derivation is that for the special case of $\rho_{XC}$ and $\rho_{ZD}$ being classical states $\rho_{XY_C}$ and $\rho_{ZY_D}$, respectively, the quantum entropies reduce to classical entropies, as defined in (\ref{eq:def-renyi-cond}). Let us additionally take the special case of the classical variables $Y_C$ and $Y_D$ encoding the same distribution, i.e.~$Y_C \equiv Y_D \equiv Y$. Then Eq.~(\ref{eq:tripartite}) reduces to
\begin{equation}
\label{eq:condition_i}
H_\alpha(X|Y) + H_\beta(Z|Y) \geq q(X,Z),
\end{equation}
which satisfies condition (i). Thus both conditions (i) and (ii) are satisfied which, as shown previously, implies a R\'enyi entropy steering inequality of the form
\begin{equation}
H_\alpha(X_B|X_A) + H_\beta(Z_B|Z_A) \geq q(X_B,Z_B),
\label{eq:ineq}
\end{equation}
with $\alpha^{-1}+\beta^{-1}=2$. A violation of this inequality excludes the existence of a local hidden state model. It is in general a sufficient, but not necessary condition for steerability. However, in the following we show two natural situations where it is also a necessary condition. We further elaborate on the tripartite EUR and the joint measurability of observables in Appendix \ref{app:implications}.

\section{Tightness for qubits and for Mutually Unbiased Bases}
\label{sec:tightness}
If Bob and Alice share a pure entangled state, steerability is equivalent to Alice using measurements which are not jointly measurable \cite{quintino_joint_2014,uola_joint_2014}. Thus the problem of determining whether Alice's measurements are jointly measurable can be examined through our entropic steering inequality (\ref{eq:ineq}). We use this connection to examine the strength of our R\'enyi entropy steering inequalities.

Consider the case of Alice using two measurements based on mutually unbiased bases (MUBs). MUBs are such that any vector of one basis has constant overlap $1/\sqrt{d}$ with any vector of the other basis.
Let us consider the case when Alice uses the computational basis $\{|j\rangle_A\}_{j=1}^d$ for her measurement $Z_A$ and its discrete Fourier transform $\{F|j\rangle_A\}_{j=1}^d$ for $X_A$, (generalizations of the Pauli $\sigma_Z$ and $\sigma_X$ in $d=2$) with
\begin{equation}
F = \frac{1}{\sqrt{d}}\sum_{j,k} \omega^{-jk} |j\rangle\langle k| \quad\text{with} \quad \omega = e^{2\pi i/d}.
\end{equation}
These measurements are not jointly measurable. In order to nudge them towards being jointly measurable we add some white noise on $Z_A$ and $X_A$, controlled by the parameters $\eta, \chi \in [0,1]$, respectively, such that both of Alice's measurement operators become white noise measurements in the limit of $\eta,\chi = 1$, as
$
Z_{A,\eta}^{j} := (1-\eta) |j\rangle\langle j| + \eta \mathbb{I}/d,
$
and similarly for $X_{A,\chi}^{i}$. The condition for joint measurability of $Z_{A,\eta}$ and $X_{A,\chi}$ reads \cite{carmeli_informationally_2012,heinosaari_invitation_2016}
\begin{equation}
\label{eq:asym_exact}
\frac{(d-1)(\eta + \chi) - \sqrt{d - (d-1)(\eta - \chi)^2}}{d-2} \leq 1.
\end{equation}

We now use our steering inequality \eqref{eq:ineq}. We first need to choose the pair of R\'enyi entropies to use, as well as the two measurements of Bob. The best choice turns out to be $\alpha = 1/2$ and $\beta = \infty$ \footnote{For $\beta=\infty$ the conditional R\'enyi entropy reduces to $H_\infty(X|Y) = H_{\text{min}}(X|Y) = \sum_y p(y) \max_x p(x|y)$ \cite{iwamoto_information_2014}.}, that is, the operationally relevant max- and min-entropies, whereas Bob's measurements are taken to be projective, in the computational basis for $Z_B$ and the Fourier-transform basis for $X_B$. Considering a maximally entangled shared state $\rho_{AB} = \sum_{ij} |ii \rangle\langle jj|$, we obtain from inequality \eqref{eq:ineq} the condition
\begin{equation}
\label{eq:asym_our}
\frac{\left(
\sqrt{\chi + \frac{1-\chi}{d}} + (d-1)\sqrt{\frac{1-\chi}{d}}\right)^2}
{1 + (d-1)\eta}
\geq 1.
\end{equation}
Despite (\ref{eq:asym_exact}) being symmetric and (\ref{eq:asym_our}) being asymmetric in $\eta$ and $\chi$, it is straightforward to show that both inequalities have equality for the same parameters $(\eta^*,\chi^*)$ for all dimensions $d$, for example by deriving constraints $\eta^*(\chi)$ from both equations and comparing them. Therefore our R\'enyi steering inequality is tight in this scenario, as it achieves the exact joint measurability bound.

\begin{figure}[t!]
\includegraphics[trim = {0 0 0 0},width=\columnwidth]{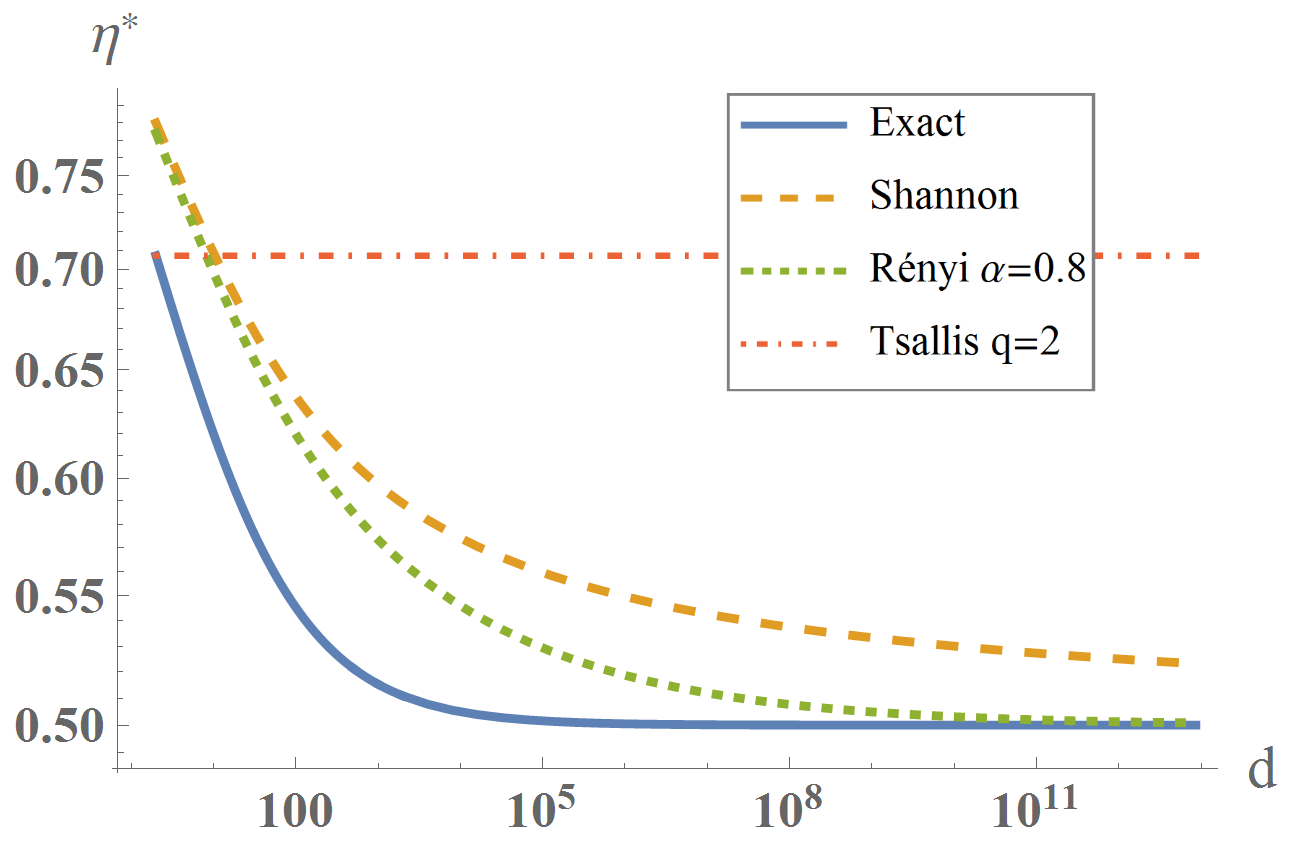}
\vspace*{-1em}
\caption{
Tolerable noise level for the detection of steering via our R\'enyi steering inequality \eqref{eq:asym_our}, for two MUBs with white noise. Here we consider the case of symmetric noise, $\eta = \chi$, as a function of the dimension $d$. The optimal noise threshold is attained for a steering inequaliy based on the min- and max-entropies ($\alpha=1/2$, $\beta = \infty$), in which case our inequality is tight. For other cases, e.g.~the Shannon entropy ($\alpha=\beta =1$) or the R\'enyi criteria with $\alpha > 1/2$, the inequality is suboptimal. For comparison the Tsallis entropy criteria derived in Ref.~\cite{costa_steering_2017} is also depicted.
}
\label{fig:alpha_08}
\end{figure}

The inequality \eqref{eq:asym_our} can be generalized for an arbitrary pair of R\'enyi entropies, with $\alpha^{-1}+\beta^{-1}=2$. One can actually show analytically, for symmetric noise $\eta = \chi$, that the derivative of the steering criteria according to $\alpha$ vanishes only at $\alpha = \beta =1 $, which proves to be a global maximum. In other words even though both the Shannon entropy-based criteria (\ref{eq:shannon-entr-ineq}) and the R\'enyi-based one (\ref{eq:ineq}) converge to the true noise threshold $\eta^* = 0.5$ in the limit of $d\rightarrow\infty$, the Shannon one performs worse than any appropriate R\'enyi entropy pair, whereas the limiting case of the max- and min-entropy pair ($\alpha=0.5, \beta = \infty$) performs best, in fact optimally. Other values of $\alpha$ seem to interpolate between the two results, getting closer to the optimal threshold more easily at higher $d$ values. See Fig.~\ref{fig:alpha_08} for an illustration.

For $d=2$ tightness holds not just for MUBs but for any two two-outcome qubit measurements. For such a generalized measurement, the POVM elements $\{Z_{A}^+, Z_{A}^- \}$ of a measurement $Z_{A}$ are described as
\begin{equation}
\label{eq:qubit_parametrization}
Z_{A}^\pm (b_z,\mathbf{z}) = \frac{\mathbb{I}\pm ( b_z \mathbb{I} + \mathbf{z}\cdot \boldsymbol\sigma)}{2},
\end{equation}
where $\mathbf{z}$ is a subnormalized Bloch vector, such that $|\mathbf{z}| = 1-\eta$, the bias parameter $b_z$ conforms to $|b_z| \leq \eta$, and  $\boldsymbol\sigma$ is a vector of the Pauli operators $\sigma_X, \sigma_Y, \sigma_Z$. The POVM $X_{A}$ has a similar form but with the bias parameter denoted as $b_x$ and a subnormalized Bloch vector $\mathbf{x}$ with a length of $|\mathbf{x}| = 1-\chi$. For such a pair of qubit binary POVMs, a necessary and sufficient condition for joint measurability has been derived in Ref.~\cite{yu_joint_2008,busch_unsharp_1986}.

To test our R\'enyi steering inequality, we use again max- and min-entropy. Consider first the case of two unbiased (i.e.~$b_z =b_x = 0 $) POVMs, and symmetric noise $\eta = \chi$. Here we show our inequality is tight analytically by choosing Bob's two measurements to be projective and MUB (i.e.~two orthogonal Pauli observables). Moreover, the average Bloch vector on Bob's side (i.e.~intermediate between the two Bloch vectors of his Pauli measurements) must coincide with the average Bloch vector for Alice's measurements (see Appendix \ref{app:renyi_elaboration}). More generally, extensive numerical analyses shows that, for arbitrary pairs of binary qubit POVMs, our R\'enyi steering inequality is tight. 

From the above examples, it is natural to ask whether the R\'enyi steering inequality is tight in general. We obtained strong evidence that this is not the case. For $d=3$, we could construct pairs of POVMs for which the inequality is not tight. Nevertheless the criterion still performs relatively well, in the sense of detecting steering for noise levels close to the joint measurability limit; see Appendix \ref{app:renyi_elaboration} for more details. More generally, for $d \geq 3$, it would be interesting to see whether there exists other sets of POVMs (besides pairs of MUBs) for which the R\'enyi steering inequality is tight. 

\section{Conclusion}
\label{sec:conclusion}
To summarize, we discussed a general connection between entropic uncertainty relations and steering. Starting from any entropic uncertainty relation satisfying two natural properties, we derived an entropy-based steering criteria, and vice versa. In particular, this allowed us to present steering inequalities based on R\'enyi entropies, which turned out to be tight in certain cases. 

We obtained the strongest steering inequality when using the min- and max-entropies. It would be interesting to understand why this asymmetry (in terms of the entropies) is beneficial for the detection of steering. Moreover, due to the operational relevance of these entropies, our steering inequality could be relevant to applications in quantum communications \cite{tomamichel_uncertainty_2011}.

Another direction concerns entropic steering criteria with more than two measurement settings. It would be interesting to see whether there are tight R\'enyi steering inequalities for an arbitrary number of measurements. Here a natural starting point would be sets of MUBs with symmetric noise, for which tight bounds for joint measurability were recently derived \cite{designolle_quantifying_2018}.

More generally, it would be interesting to see if the connection between entropic uncertainty relations and steering can be made even more general. 
In Appendix \ref{app:weaker_conditions} we give an alternative derivation of our result based on two other natural properties of the conditional entropy (uncorrelated limit and concavity). An interesting question is whether our method can be applied for Tsallis entropies. It turns out to be not so straightforward; some aspects of this are discussed in Appendix \ref{app:tsallis}.

Finally, our methods also allow for the construction of steering inequalities for continuous variable systems, and it would be of interest to investigate their properties and potential practical relevance.

\begin{acknowledgments}
We would like to thank S. Designolle, P. Skrzypczyk, and R. Uola for discussions. We acknowledge financial support from the Swiss National Science Foundation (Starting grant DIAQ, and QSIT), and the European Research Council (ERC MEC).
\end{acknowledgments}


\appendix
\section{Alternative requirements}
\label{app:weaker_conditions}
The question arises whether condition (i) can be derived from native properties of the entropies involved. Here we demonstrate that two simple and natural properties of entropy are enough to derive an EUR conditioned on classical side-information, from an EUR.

Let $\{\tilde H_k(\cdot|\cdot)_\rho\}_k$ be a family of classical conditional entropies, evaluated on the statistics of measurement performed on the quantum system $\rho$. Consider the following two properties of conditional entropies.
\begin{itemize}
\item[(iii)] \emph{Uncorrelated limit} For any product state $\rho = \rho_{B}\otimes \rho_{A}$ and any measurements $X_B,X_A$
\[\tilde H(X_B|X_A)_\rho = \tilde H(X_B)_\rho.\]
\item[(iv)] \emph{Concavity of conditional entropy} For any state $\rho$ which is a convex combination of some states $\rho^k$, with weights $p(k)$, i.e.~$\rho = \sum_k p(k) \rho^k$, it holds that
\[\tilde H(X_B|X_A)_{\rho } \geq \sum_k p(k) \tilde H(X_B|X_A)_{\rho^k }.\]
\end{itemize}

\emph{Claim.} Given an EUR of the form
\begin{equation}
\label{eq:appendixEUR}
\sum_k \tilde H_k(X_{B,k})_\rho\geq \tilde q(\{X_{B,k}\}),
\end{equation}
if each $\tilde H_k$ entropy satisfies properties (iii) and (iv), then the following inequality is satisfied for any local hidden state model,
\begin{equation}
\sum_k \tilde H_k(X_{B,k}|X_{A,k})_\rho\geq \tilde q(\{X_{B,k}\}), 
\end{equation}
and hence certifies steering if it is violated.

\emph{Proof.} Consider a local hidden state model, i.e.~a situation in which the joint state of Alice and Bob is of the form (\ref{eq:clqustate}), i.e.
\begin{equation}
\label{eq:tempeditemptemp}
\rho_{AB} = \sum_{\lambda} p(\lambda) \ket{\lambda} \bra{\lambda}_A \otimes \sigma^{\lambda}_B.
\end{equation}
Each component of such a mixture, $\ket{\lambda} \bra{\lambda}_A \otimes \sigma^{\lambda}_B$, which for simplicity we will denote as $\rho^\lambda$, has a product state structure as in condition (iii), which implies that for any $\lambda, k$ and any $X_{B,k}$ and $X_{A,k}$ measurements
\begin{equation}
\label{eq:appendixtemp1}
\tilde H_k(X_{B,k}|X_{A,k})_{\rho^\lambda} = \tilde H_k(X_{B,k})_{\rho^\lambda}.
\end{equation}
Upon summing over all $k$ values that appear in the EUR (\ref{eq:appendixEUR}), the right hand side can be lower bounded by $q(\{X_{B,k}\})$ as
\begin{equation}
\sum_k \tilde H_k(X_{B,k})_{\rho^\lambda} \geq \tilde q(\{X_{B,k}\}),
\end{equation}
hence, through Eq.~(\ref{eq:appendixtemp1}), we arrive at
\begin{equation}
\sum_k \tilde H_k(X_{B,k}|X_{A,k})_{\rho^\lambda} \geq \tilde q(\{X_{B,k}\}).
\end{equation}
Let us now multiply both sides by $p(\lambda)$ and sum over all values of $\lambda$. Then by property (iv), the left hand side can be upper bounded as
\begin{equation}
\sum_k \tilde H_k(X_{B,k}|X_{A,k})_{\rho} \geq \sum_{\lambda,k} p(\lambda) \tilde H_k(X_{B,k}|X_{A,k})_{\rho^\lambda},
\end{equation}
leading us to
\begin{equation}
\label{eq:temptemp}
\sum_k \tilde H_k(X_{B,k}|X_{A,k})_{\rho} \geq \tilde q(\{X_{B,k}\}).
\end{equation}
Recall that the structure of $\rho_{AB}$ is given by the local hidden state model, i.e.~by Eq.~(\ref{eq:tempeditemptemp}). Thus we arrive at the claim. $\hfill \square$

Note that for some entropy functions, such as the Shannon entropy, property (iv) follows from (ii), hence properties \{(ii), (iii)\} are enough for the proof. It is straightforward to see this from the definition of the conditional Shannon entropy,
\begin{equation}
H(X_B|K) = \sum_k p(k) H(X_B|K=k).
\end{equation}
Treating the label $K$ as a random variable one arrives at property (iv) through a single application of property (ii),
\begin{equation}
H(X_B|X_A)_{\rho} \geq \sum_k p(k) H(X_B|X_A, K=k)_{\rho^k}.
\end{equation}
For more examples of conditional entropies which are concave, see \cite{gigena_generalized_2014}.

Note that in the derivation shown here, if one uses any classical random variable $Y$ in Eq.~(\ref{eq:appendixtemp1}) instead of the $X_{A,k}$ random variables, then one immediately derives (i) from \{(iii), (iv)\}. One might think, \emph{a priori}, that then (ii) is also necessary to continue the proof, as performed in the main text. Surprisingly this is not the case, since properties \{(iii), (iv)\} are sufficient, as shown here.

As a final comment, note that in the proof given here in the appendix, it is also not possible to use a state-dependent bound, since then the $\sum_\lambda p(\lambda)$ would not factor out in (\ref{eq:temptemp}).

\section{Application of the R\'enyi steering criteria}
\label{app:renyi_elaboration}

\begin{figure}[t]
\includegraphics[width=0.6\columnwidth]{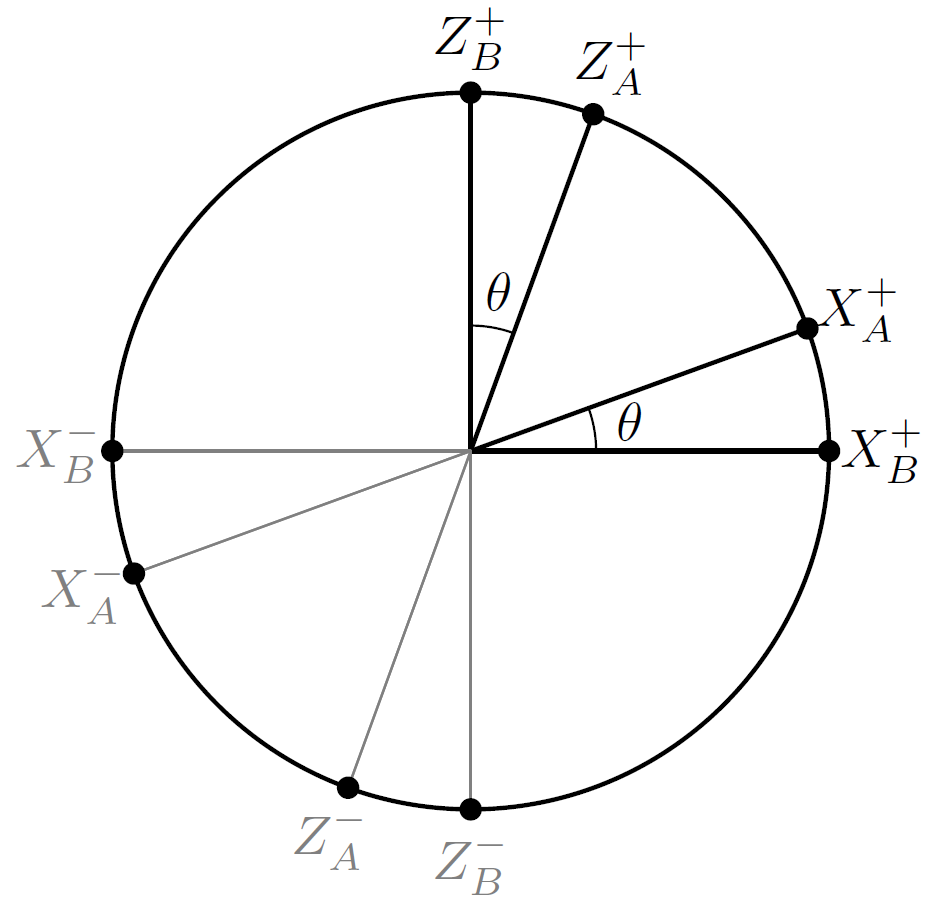}
\vspace*{-1em}
\caption{
A section of the Bloch sphere designated by Alice's measurements ($Z_A^\pm,X_A^\pm$). Optimal measurements of Bob ($Z_B^\pm,X_B^\pm$) are shown for the case of Alice's measurements having symmetric noise ($\eta = \chi \in [0,1]$), no bias ($b_z=b_x=0$) and for Alice and Bob sharing a maximally entangled state $\frac{1}{\sqrt{2}} (\ket{00}+\ket{11})$.
}
\label{fig:qubit_angles}
\end{figure}

Here we give more insight into the joint measurability noise threshold results achieved using the R\'enyi entropy based steering criteria (\ref{eq:ineq}). First we show analytically for $d=2$ which measurements are optimal for Bob to detect steering when Alice's measurements have symmetric noise on them. For a full study of the parameter space $\{\eta,\chi,b_z,b_x\}$, according to Eq.~\eqref{eq:qubit_parametrization}, we rely on numerical analysis. After finishing the discussion of the $d=2$ case, we proceed to show a family of measurements in $d=3$ which are not MUBs, for which the R\'enyi criteria is not tight, but is still close to the true noise boundary.

For $d=2$ we now consider the case when Alice's measurements $Z_A$ and $X_A$ have no bias, $b_z = b_x =0$, and have the same amount of noise on them, $\eta = \chi$ (see Eq.~\eqref{eq:qubit_parametrization} for parametrization of measurements). In such a scenario Alice's measurements can be represented by two normalized vectors in the Bloch sphere, one for her $Z_A^+$ and one for her $X_A^+$ POVM element. The goal is to find the noise parameter $\eta^*$ for which these two measurements become jointly measurable. The two vectors single out a plane in the Bloch sphere. We show that it is in this plane in which Bob can find his optimal measurements. Namely his measurements' eigenvectors should be orthogonal in the Bloch sphere such that they are symmetrically surrounding Alice's measurement vectors, as shown in Fig.~\ref{fig:qubit_angles}. In this way Bob's measurements are two MUBs. For such a measurement setup, the conditional probabilities for a given noise $\eta$ and angle $\theta$ are $\frac{1}{2}(1\pm\eta \cos \theta)$, the sign depending on whether the eigenvectors are in the same hemisphere of the Bloch sphere ($+$), e.g.~$Z_A^+,Z_B^+$ or not ($-$), e.g.~$Z_A^+,Z_B^-$. The noise threshold from the R\'enyi criteria follows from straightforward substitution into (\ref{eq:ineq}), as 
\begin{equation}
\eta^*_R = \frac{1}{\sqrt{2} \cos \theta}.
\label{eq:renyi_qubit}
\end{equation}
The exact noise threshold for such a scenario is \cite{busch_unsharp_1986}
\begin{align}
\eta^* =& \frac{2}{\|z_A^+ + x_A^+\|+\|z_A^+ - x_A^+\|} = \\
&\frac{\sqrt{2}}{\sqrt{1-\sin(2\theta)} + \sqrt{1+\sin(2\theta)}},
\label{eq:exact_qubit}
\end{align}
where the $z_A^+,x_A^+$ vectors now explicitly denote the Bloch sphere vectors of $Z_A^+$ and $X_A^+$, respectively. Equations (\ref{eq:renyi_qubit}) and (\ref{eq:exact_qubit}) can be shown to be equivalent through the use of trigonometric equality $\frac{1}{2} (1+\cos2\theta) = \cos^2\theta$.
\begin{figure}[t]
\includegraphics[width=0.85\columnwidth]{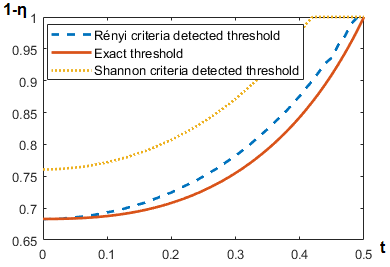}
\vspace*{-1em}
\caption{
We depict sharpness, i.e.~$1-\eta$, as a function of the parameter $t$ in $d=3$. For $t=0$ we are in the MUB case, for which we have previously shown tightness of the R\'enyi criteria, hence the exact threshold and the detected threshold are the same. For $t=0.5$ Alice's measurements overlap, hence even for $\eta=0$ they are jointly measurable. Although the R\'enyi criteria are not tight, they outperforms the Shannon criteria.
}
\label{fig:d3}
\end{figure}

For dimensions larger than two, we created a natural generalization of the rotation parametrized by $\theta$ in $d=2$, illustrated in Fig.~\ref{fig:qubit_angles}. In $d=3$, apart from the two Fourier connected bases, we take an element $\ket{Y_A,1}$ of a third MUB $Y_A$. Then for $t\in[0,0.5]$ and $i\in\{1,2,3\}$, the basis elements $\ket{Z_{A,i}^{t}}$ of $Z_A^t$ are
\begin{equation}
\ket{Z_{A,i}^t} = \exp \left(-t\frac{2\pi i}{3} \ket{Y_{A,1}}\bra{Y_{A,1}}\right) \ket{Z_{A,i}^0},
\end{equation}
and similarly for the basis elements $\ket{X_{A,i}^{t}}$ of the basis $X_{A}^{t}$. As mentioned before, we choose $X_{A}^{0}$ and $Z_{A}^{0}$ to be the Fourier connected bases. As $t$ increases, so does the overlap between the elements of the bases $X_{A}^{t}$ and $Z_{A}^{t}$, until they become identical for $t=0.5$. For such a family of measurements, for $i\neq j$ the absolute overlaps $|\braket{Z_{A,i}^t}{X_{A,j}^t}|$ are all equal, and similarly for $i=j$. For such a family of measurements the true noise thresholds as well as the thresholds detected by the R\'enyi criteria can be seen in Fig.~\ref{fig:d3}.

The question remains open whether there exists a family of measurements in dimension $d>2$, which are not MUBs, such that the R\'enyi criteria can detect their joint measurability tightly. It would be reasonable to examine measurements which achieve equality of the considered EUR, e.g. measurements discussed in Ref.~\cite{abdelkhalek_optimality_2015}.

\section{Tsallis entropy}
\label{app:tsallis}

Steering inequalities using the Tsallis entropy have been derived in Ref.~\cite{costa_steering_2017}, using principles different than the ones shown here, relying on properties such as the joint convexity of the relative entropy. Here we examine the question of whether Tsallis entropy results can be derived using our methods.

The Tsallis $q$-entropy, defined as \cite{furuichi_information_2006}
\begin{equation}
S_q(X) = -\sum_x p(x)^q \ln_q(p(x)),
\end{equation}
for $q\in\mathbb{R}^+\backslash \{1\}$ and $\ln_q(p(x)) := (p(x)^{1-q}-1)/(1-q)$, is a quantity that is oftentimes used in nonextensive statistical physics. For a generic $q$, it not an additive entropy. However, it does have a pseudo-additive property. That is, for $X$ and $Y$ being independent random variables,
\begin{equation}
\label{subadditivity_long}
S_q(X,Y) = S_q(X) + S_q(Y) + (1-q)S_q(X)S_q(Y).
\end{equation}
In this way one can interpret the parameter $q$ as being a measure of how non-additive the entropy is. In the limit of $q\rightarrow 1$ we recover the Shannon entropy, including its additivity.

For the conditional Tsallis $q$-entropy, defined as
\begin{equation}
S_q(X|Y) = \sum_y p(y)^q S_q(X|Y=y),
\end{equation}
the chain rule holds,
\begin{equation}
S_q(X,Y) = S_q(X|Y) + S_q(Y),
\end{equation}
as well as conditon (ii) for $q\geq 1$, i.e.
\begin{equation}
S_q(X|Y) \leq S_q(X),
\end{equation}
with equality if and only if $q=1$ and $X$ and $Y$ being independent random variables. Subadditivity for independent random variables and $q\geq 1$ follows straightforward from the pseudo-additive property (\ref{subadditivity_long}), however, subadditivity can be extended to generic random variables through the use of the ''conditioning reduced entropy'' (condition (ii)) property, i.e.~
\begin{equation}
S_q(X,Y) \leq S_q(X) + S_q(Y),
\end{equation}
with equality only for $q=1$ and $X$ and $Y$ being independent. 

One would hope that one could derive condition (i) starting from a Tsallis EUR, such as \cite{rastegin_uncertainty_2013}
\begin{equation}
\sum_k S_q(X_{B,k}) \geq C_B^{(q)}(m),
\end{equation}
where $C_B^{(q)}(m) = m \ln_q\left( \frac{md}{d+m-1}\right)$ for a set of $m$ MUBs in dimension $d$, and $q\in(0,2]$. In Appendix \ref{app:weaker_conditions}, we could derive (i) either from conditions \{(iii), (iv)\} or from \{(ii), (iii)\} (for entropies with a specific property). Both derivation rely on condition (iii). Unfortunately, for Tsallis entropies with $q\geq1$, condition (iii) does not hold, i.e.~for measurements $X$ and $Y$ executed on a product state $\rho_{AB}=\rho_A\otimes\rho_B$, in general
\begin{equation}
S_q(X|Y) \neq S_q(X).
\end{equation}
This is easy to see by applying the chain rule and then subadditivity to $S_q(X|Y)$, that is
\begin{equation}
S_q(X|Y) = S_q(X,Y) - S_q(Y) \leq S_q(X),
\end{equation}
with equality \emph{only} for $q=1$ and $p(x|y) = p(x)$ for all $x$ and $y$. That is, property (iii) only holds for the limiting case of the Shannon entropy.

This line of thought demonstrates that the Tsallis entropy is not a natural fit for our derivation. Another derivation was applied in Ref.~\cite{costa_steering_2017}, which takes into account the nonadditvity of the Tsallis entropy and results in the steering inequality (for $m$ MUBs in dimension $d$)
\begin{equation}
\sum_k S_q(X_{B,k}|X_{A,k}) + (1-q)C(X_{A,k},X_{B,k}) \geq  C_B^{(q)}(m),
\end{equation}
with the correction term $C(X_{A,k},X_{B,k})$ (defined in Ref.~\cite{costa_steering_2017}). These Tsallis entropy based steering criteria seem to be less strong than the R\'enyi ones derived here for two MUBs, as illustrated in Fig.~\ref{fig:alpha_08}. However, the Tsallis result is easily extendible to more measurements.

\section{Implications for joint measurability}
\label{app:implications}

We will now show that by deriving (\ref{eq:ineq}) we have shown a non-trivial reduction of the tripartite quantum side-information EUR (\ref{eq:tripartite}), which we rewrite here with slightly different notation
\begin{equation}
\label{eq:tripartite_new}
H_\alpha^Q(X_B|C')_{\rho} + H_\beta^Q(Z_B|D')_{\rho} \geq q(X_B,Z_B),
\end{equation}
for $\alpha,\beta \geq 1/2$ and $\alpha^{-1}+\beta^{-1}=2$.
We used the realization that if measurements $X_{CD}$ and $Z_{CD}$ (acting on some $\rho_{CD}$) are jointly measurable, then their joint results can be stored as classical information, which in turn can be copied and encoded into quantum systems $C'$ and $D'$ of appropriate size. Thus for $X_{CD}$, $Z_{CD}$ being jointly measurable it holds that
\begin{equation}
\label{eq:tripartite_new_reduced_JM}
H_\alpha(X_B|X_{CD}) + H_\beta(Z_B|Z_{CD}) \geq q(X_B,Z_B).
\end{equation}
The specific case of $X_{CD} = X_C \otimes \mathbb{I}_D$ and $Z_{CD} = \mathbb{I}_C \otimes Z_D$ allows us to reduce (\ref{eq:tripartite_new_reduced_JM}) to
\begin{equation}
\label{eq:tripartite_new_reduced_separated}
H_\alpha(X_B|X_C) + H_\beta(Z_B|Z_D) \geq q(X_B,Z_B).
\end{equation}

Note that by the data processing inequality, Eq.~(\ref{eq:tripartite_new_reduced_separated}) follows directly from (\ref{eq:tripartite_new}) by measuring $X_C$ on system $C'$ and $Z_D$ and system $D'$, separately. This is the most operationally natural reduction of (\ref{eq:tripartite_new}). Yet here we have introduced a nontrivial intermediate step (\ref{eq:tripartite_new_reduced_JM}). Note that although (\ref{eq:tripartite_new}) and (\ref{eq:tripartite_new_reduced_separated}) are both valid for conditioning on (results from) \emph{separate systems}, the intermediate step (\ref{eq:tripartite_new_reduced_JM}) conditions on results from a \emph{joint system}, but with the additional constraint that the measurements are jointly measurable. This sparks the intuition that as long as we use jointly measurable measurements on a quantum system, the system can effectively be treated as a bipartite system.

This view on jointly measurable measurements, originating from the connection of joint measurability to steering, could open new avenues for understanding, interpreting and deriving relations for measurements, in particular ones using conditional entropies.


\end{document}